\documentstyle[emulateapj,psfig]{article}

\received{August 20, 1999}
\accepted{December 1, 1999}
\slugcomment{(Published in {\it The Astrophysical Journal Letters.
Received August 20, 1999. Accepted December 1, 1999.})}

\begin{document}

\title{Evidence for Multiple Mergers among Ultraluminous IR Galaxies
(ULIRG{\lowercase{s}}): Remnants of Compact Groups? \footnote{
Based on observations with the NASA/ESA Hubble Space Telescope, obtained at
the Space Telescope Science Institute, which is operated by the Association of
Universities for Research in Astronomy, Inc. under NASA contract No.
NAS5-26555.}
}


\author{Kirk D. Borne}
\affil{Raytheon Information Technology and Scientific Services,
Code 631, NASA Goddard Space Flight Center, Greenbelt, MD 20771;
{\tt{borne@rings.gsfc.nasa.gov}}
}

\author{Howard Bushouse and Ray A. Lucas}
\affil{Space Telescope Science Institute, 3700 San Martin Drive, 
Baltimore, MD 21218;
{\tt{bushouse@stsci.edu, lucas@stsci.edu }}
}

\and

\author{Luis Colina}
\affil{Instituto de Fisica de Cantabria (CSIC-UC),
Facultad de Ciencias, 39005 Santander, Spain;
{\tt{colina@ifca.unican.es}}
}

\begin{abstract}
In a large sample of ULIRGs
imaged with {\it{HST}}, we have identified a significant subsample 
that shows evidence for multiple mergers.  
The evidence is seen among two classes of ULIRGs: 
(1)~those with multiple remnant nuclei in their core, 
sometimes accompanied by a complex system of tidal tails; and
(2)~those that are in fact dense groupings of interacting
(soon-to-merge) galaxies.  We conservatively estimate that,
in the redshift range 0.05$<$$z$$<$0.20, at least
20 (out of 99) ULIRGs satisfy one or both of these criteria.
We present several cases and discuss the possibility that 
the progenitors of ULIRGs may be the
more classical weakly interacting compact groups of galaxies (Hickson).
An evolutionary progression is consistent with the results: 
from compact groups to pairs to ULIRGs to elliptical galaxies.
The last step follows the blowout of gas and dust from the ULIRG.  
\end{abstract}



\section{Introduction} \label{introd}

In the 1980's, two samples of galaxies were identified 
that subsequently became the subjects of vigorous research
activity.  One of these was the sample of compact groups 
identified by Hickson (1982; see Hickson~1997 for a
review; hereafter, we call these CGs).  The other was the
sample of ultraluminous IR galaxies (ULIRGs) identified in the
IRAS all-sky survey (Sanders {\it{et al.}}~1988; 
see Sanders \& Mirabel 1996 for a review).  
The high IR luminosity ($L$[8--1000$\mu$m] $> 10^{12}L_\odot$) 
of ULIRGs is commonly thought to arise from dust absorption and
IR re-emission of the intense but obscured starburst+AGN
radiation field.  While it was known early on that
ULIRGs nearly always show evidence for interactions
(collisions/mergers), many investigators have debated the interaction
fraction among ULIRGs.  Published values range from $\sim$100\
(Duc, Mirabel, \& Maza 1997; 
Borne {\it{et al.}}~1999a, 1999b) 
down to $\sim$50\
(Auriere {\it{et al.}}~1996).
Under the assumption that mergers of gas-rich galaxies 
stimulate the ULIRG phase,
several theoretical studies have modeled the dynamical events leading to the
ultra-starburst event 
(Mihos \& Hernquist 1996, and references therein; hereafter MH).  
In spite of this intense research activity,
scant attention has been given to
identifying the complete set of progenitor systems that
may produce the ULIRGs seen in the local universe.
While there are numerous examples of merger remnants and strongly
interacting pairs of galaxies at low redshift,   
there is only one {\it{bona fide}}
ULIRG with $cz<10^4$ km s$^{-1}$ (Arp~220).
Hence it is unreasonable to believe that merging {\it{pairs}}
of spirals alone can account for the dynamical diversity of the 
ULIRG population.  There must be some other class of progenitors.  
As hypothesized by Xia {\it{et al.}}~(1999),
a possible progenitor sample is the set of CGs.

We undertake, in \S \ref{commonalities},
a comparative analysis of ULIRG and CG properties in the light
of several new results.  We then present 
a short description of our observations 
in \S \ref{hstobs}, our sample of multiple--merger candidates 
in \S \ref{multcands}, and a discussion of a possible CG--ULIRG 
connection in \S \ref{discussion}.

\section{ULIRGs and Compact Groups: Common Features} \label{commonalities}

One of the remarkable attributes of ULIRGs is their unique 
``once-in-a-galactic-lifetime'' IR-ultraluminous status (MH).  
A galaxy 
may enjoy this 
status only once  
since the very property that defines the high-IR
luminosity phase (i.e., the intense dust-absorbed and re-radiated 
radiation field) will likely blow out the dust from the galaxy.  
Furthermore, the intense starburst phase that characterizes 
the overwhelming majority of ULIRGs (Genzel {\it{et al.}}~1998)
will likely consume the available gas supply
or otherwise render the gas unavailable for future star formation
(MH).  
These physical processes 
consequently prevent the onset of a subsequent
dust-obscured super-starburst phase.  The ULIRG phase is thus transitory, 
with a somewhat uncertain life span (duty cycle) 
for detectability and classification as a
ULIRG.  The duration of the phase is probably
$\sim$$10^8$ yr (Devriendt, Guiderdoni, \& Sadat 1999).
As a result, ULIRGs are identified 
at a special phase in their dynamical history.  

CGs may also be caught in a special state. 
Their dynamical timescales were originally 
considered to be quite short (0.01--0.1 $H_0^{-1}$), 
implying a strong merging instability (Barnes 1985, 1989; Mamon 1987).
Subsequent simulations that embedded the CG within a common massive halo
significantly increased the merger timescale to $\sim$$H_0^{-1}$ 
(Governato, Tozzi, \& Cavaliere 1996; 
Athanassoula, Makino, \& Bosma 1997).  
The common massive halo is consistent with 
current hierarchical merging scenarios (Kolatt {\it{et al.}}~1999). 
If the actual merging timescales are somewhere
between these extremes, then CGs (like ULIRGs)
are also transitory and will ultimately evolve out of the CG sample through
the merger and coalescence of their constituent group members,
as indicated in the 
numerical investigations of multiple mergers within a CG setting
by Weil \& Hernquist (1994, 1996).

Xia {\it{et al.}}~(1999)
suggested that ULIRGs are the
dynamical descendents of CGs.  The implied multiple--merger
scenario was investigated in the case of Arp~220 by
Taniguchi \& Shioya (1998), who proposed multiple mergers as the origin
for most ULIRGs.
To test whether ``evolved CGs'' become ULIRGs,
we compare the relative space densities
and dynamical ages of ULIRGs and CGs.
For a redshift survey-selected CG sample, Barton {\it{et al.}}~(1996) 
give $\Phi = 1.4 \times 10^{-4} h^3 = 4.8 \times 10^{-5}$ Mpc$^{-3}$.
(For this paper, we assume $H_0$ = 100$h$ km s$^{-1}$ Mpc$^{-1}$, $h$=0.7,
and $q_o = {1 \over 2}$.)
For the IRAS 1-Jy ULIRG sample, Kim \& Sanders (1998)
give $\Phi = 1.8 \times 10^{-7} (h/75)^3 = 1.5 \times 10^{-7}$ Mpc$^{-3}$
(summed over all luminosity bins),
with strong evolution proportional to $(1+z)^n$,
where $n = 7.6 \pm 3.2$.  Taken literally, the ratio
of space densities is $\Phi$(CGs):$\Phi$(ULIRGs) = 32:1,
similar to the ratio of life spans for the two 
populations (few Gyr : 100 Myr), assuming the longer CG dynamical age.
This is therefore consistent with 
the notion that ULIRGs evolve out of the CG
population.  Of course, more detailed plausibility arguments are required
to validate such a notion, such as comparing 
the gas content, 
X-ray and IR emission properties,
galaxy morphological mix, 
total mass, and 
the wider environment of CGs 
to the corresponding properties of ULIRGs.  
Further investigations along these lines would likely be very illuminating.

Given the potentially short life spans of both ULIRGs
and CGs, it is curious that we see very many of either.  
The best interpretation
of this is that these are the surviving members (the tail of the
distribution) of evolving populations.  CGs may be continuously
replenished through the dynamical evolution of loose groups
(Diaferio, Geller, \& Ramella 1994; Ramella {\it{et al.}}~1994), 
with the ones that we see today being the tail
of the hierarchical evolution of large-scale structure,
which favors the formation of small-scale structures
early in the Universe.  Similarly, 
what few ULIRGs we see at low redshift are likely the tail of a previously
rich distribution of ULIRGs, as evidenced by their strong 
redshift evolution (Kim \& Sanders 1998).  Given the 
above discussions, it is thus not surprising that:
(a)~a significant population of ultraluminous dusty starbursting 
galaxies at high redshifts ($z$$\sim$1--5, peaking at $z$$<$3) 
has been identified as counterparts of the 
Submillimeter Common-User Bolometer Array
(SCUBA) submm sources  
and as the contributors to the IR background (Dwek {\it{et al.}}~1998;
Hughes {\it{et al.}}~1998; Barger {\it{et al.}}~1998; 
Smail {\it{et al.}}~1998; Blain {\it{et al.}}~1999; 
Trentham, Blain, \& Goldader 1999);
and that (b)~this same epoch ($z$$\sim$3) witnesses the hierarchical 
merging of dense configurations of sub-halo
galactic-scale fragments within massive halos 
(Somerville, Primack, \& Faber 1998; Kolatt {\it{et al.}}~1999).  Such
fragments within massive halos 
are consistent with recent theoretical models of CGs
(Athanassoula {\it{et al.}}~1997), 
and the ultraluminous dusty starburst
SCUBA sources are consistent with ULIRGs --- the connection between
ULIRGs and CGs thus seems nearly certain in the cosmological setting.

\section{{\it{Hubble Space Telescope}} Imaging Observations} \label{hstobs}
 
{\it{Hubble Space Telescope}} ({\it{HST}}) images were obtained 
with the Wide Field Planetary Camera 2
(WFPC2) camera in the F814W $I$-band filter for a large sample
of ULIRGs (Borne {\it{et al.}}~1997a, 1997b, 1999a, 1999b).  
For each target in our survey, we obtained two 400~s images 
in order to remove the effects of cosmic-ray radiation events in the CCDs.
The angular scale is 0.0996$''$ per pixel (Trauger {\it{et al.}}~1994).
Our comprehensive
WFPC2 Snapshot Atlas paper (Borne {\it{et al.}}~2000,
in preparation) will provide 
a thorough description of each of the 
more than 120 ULIRGs in our {\it{HST}} survey.

\section{Multiple-Merger Candidates} \label{multcands}
 
\begin{figure} 
\par\noindent
\centerline{\psfig{figure=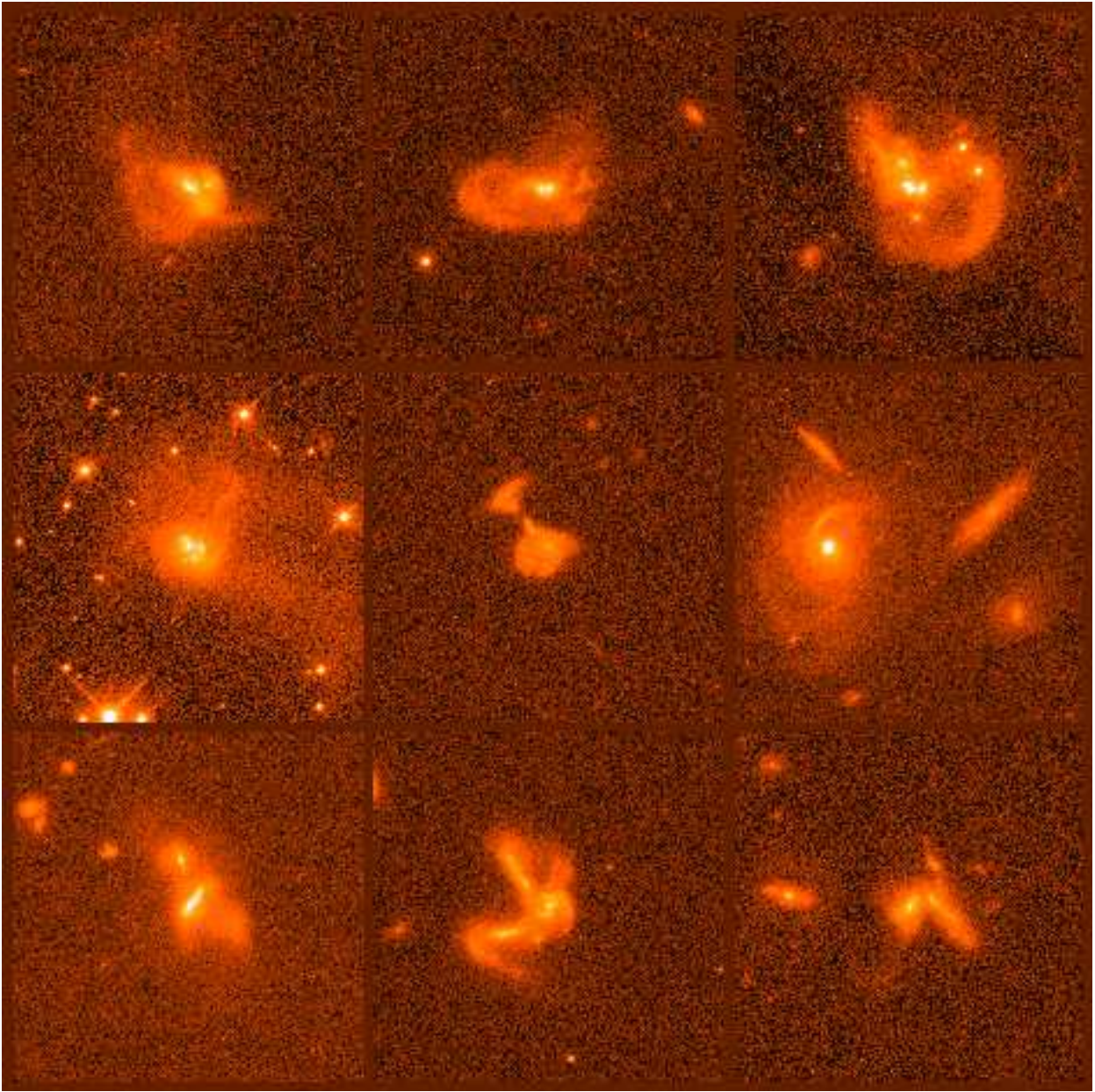,height=16.5cm,width=16.5cm}}
\caption{
{\it{HST}} WFPC2 $I$-band (F814W filter) images of a
sample of ULIRGs whose morphologies appear to be derived from $\ge 2$ mergers.
These are listed in Table~\ref{tab:candidates}, ordered by
their location in this figure: left to right for the top row
({\it{HST}} visit numbers B7, C3, and 44), for the middle row
(visits 15, E1, and 25), and for the bottom row
(visits 27, 54, and F5).  Each image is 30$''$$\times$30$''$.
}
\label{fig:candidates}
\end{figure}

With the high angular resolution ($\sim$0.1--0.2$''$) of {\it{HST}}, some
ULIRGs that were previously classified as ``non-interacting'' now 
show secondary nuclei at their centers and additional tidal
features (see examples in Borne {\it{et al.}}~1999a, 1999b).  It thus appears
that the fraction of ULIRGs showing evidence for interaction is very
nearly 100\%.  
Our {\it{HST}} images indicate in some cases that the mergers
are well developed with single nuclei and full coalescence, while
others show clear evidence for two or more nuclei, while the rest
of the sample can best be described as compact groupings of $\ge$2
distinct galaxies.  It is not obvious from this that there is a well-defined
point during such interactions at which the ULIRG phase develops, nor
is it clear what the duration of the ultraluminous phase should be.
One possible explanation for this {\it{dynamical diversity}} is the
multiple--merger model.  In this scenario, the existence of double
active galactic nuclei/starburst
nuclei is taken as evidence of more than one merger, following the creation
of the current starburst nuclei from a prior set of mergers
(Taniguchi \& Shioya 1998) --- 
the currently observed merger would be at least
the third merger in the evolutionary sequence.
Figure~\ref{fig:candidates} presents images of nine ULIRGs from our {\it{HST}}
sample that appear to have evolved from multiple mergers.  The evidence
for this includes the presence of either more than two distinct and well
separated remnant nuclei or more than two
component galaxies, often with an unusually complex system of tidal tails 
and loops that suggest multiple dynamical origins of the tidal features,
as seen in the simulations of Weil \& Hernquist (1994, 1996).
A connection between the ULIRG and CG populations 
is therefore supported by these {\it{HST}} observations.
Table~\ref{tab:candidates} lists our
multiple-merger ULIRG candidates.  For a nearly complete subsample of 
99 ULIRGs in the redshift range $0.05<z<0.20$, we identify at least 20  
either as on-going mergers in a group ($N_{\rm{gal}} > 2$) 
or as remnants of multiple ($\ge$2) mergers.

The most serious concern with the multiple-merger hypothesis
is the ubiquitous presence of multiple condensations (in optical
images of ULIRGs) that emerge through a complex dust obscuration pattern.
To minimize this effect, cases with multiple knots in the core were
specifically excluded.  Only those cases that clearly reveal separate
optically luminous galactic components were selected.
In addition, objects were selected if they had very
complex tidal features (several tails), which may indicate multiple
mergers (independent of any core condensations or nuclear dust obscuration).
These selection criteria reduced the fraction of
multiple-merger candidates to only 20\% of ULIRGs.  If questionable
cases with multiple knots were also included, then the ``multiple-merger
fraction'' would be more like 80\%.  It is imperative that radio and IR
imaging be obtained in order to test the multiple-merger classification
for the objects in our sample (Table~\ref{tab:candidates}).

Another ``feature'' of some ULIRGs is that they
are in that category simply because several galaxies
appear in the large IRAS beam and thus
conspire to produce a high IR flux.  
No single object or interacting pair
in those systems is really a ULIRG.  
It is expected that some higher-redshift CGs
(compared to local CGs) would occupy one IRAS beam and hence be 
classified as ULIRGs.  These particular systems may be 
``once and future ULIRGs''.

\section{Discussion: The CG--ULIRG Connection} \label{discussion}

We have used the {\it{HST}} to study a large sample of ULIRGs. 
The images are consistent with a multiple--merger
origin for a significant fraction of the sample,
whose rich variety of morphologies almost certainly
relates to diverse interaction histories.  However, morphology
alone cannot confirm the hypothesis.  We are therefore
conducting a detailed photometric and surface brightness
analysis of nearly 30 ULIRGs with both I-band and H-band ({\it{HST}} NICMOS)
imaging to test whether the nuclei that we see are in fact
galactic nuclei or super-starburst knots (Colina {\it{et al.}}~2000,
in preparation; Bushouse {\it{et al.}}~2000, in preparation).  
Our preliminary results confirm that the observed cores are galactic nuclei.

ULIRGs and CGs may share a common evolutionary path.
We find two morphological classes among our ULIRGs
that are consistent with a multiple--merger origin and hence 
support the hypothesis that CGs
are the progenitors for some ULIRGs.  These classes are:
(1)~ULIRGs with multiple remnant nuclei in their core, 
sometimes accompanied by a complex system of tidal tails; and
(2)~ULIRGs that are in fact dense groupings of interacting
(eventually merging) galaxies.
These classes are assigned in Table~\ref{tab:candidates}.
Borne {\it{et al.}}~(1999b) find an equal likelihood 
for a ULIRG to be a recent merger (single) as to be involved in an 
on-going collision (multiple) and find 
very little variation in the mean $L_{\rm{IR}}$ between 
these two categories.  This would be consistent with a series
of mergers taking place in a typical ULIRG, producing a sustained
super-starburst --- the timing of each burst is
strongly determined by the orbital orientation (prograde or
retrograde) and the internal structure of the
merging galaxies (bulge or no-bulge), as investigated in
detail by MH.  Consequently, deducing 
the phase of interaction for individual ULIRGs will be quite
complicated: Which merger are we now witnessing?  
Is it the first, the second, or the N-th?
While our observations conclusively show that some ULIRGs are the result
of multiple mergers,  
more observational and theoretical investigations are required 
to validate the multiple--merger model for the sample, 
to estimate better the multiple--merger fraction,
to verify a CG--ULIRG connection, and 
to elucidate the dynamical origin, state, and fate of these remarkable objects.

\acknowledgements
 
Support for this work was provided by NASA through 
grant number GO-06346.01-95A from the Space
Telescope Science Institute, which is operated by 
AURA, Inc., under NASA contract NAS5-26555.
K.D.B.~thanks Raytheon for providing financial support
during his Sabbatical Leave
and thanks the STScI for sponsoring his Sabbatical Visit.
We thank our STScI program coordinator Andy Lubenow
and contact scientist Keith Noll for assistance
in implementing the {\it{HST}} program, and we thank
the referee for helpful comments.


\footnotesize
\begin{deluxetable}{clrrlcc}
\tablewidth{450pt}
\tablecaption{ULIRG Multiple-Merger Candidates
\label{tab:candidates} }
\tablehead{
\colhead{{\it{HST}} Visit \#} &
\colhead{IRAS Name} &
\colhead{RA(J2000)} &
\colhead{Dec(J2000)} &
\colhead{Morphology} &
\colhead{Redshift} &
\colhead{$\log L_{60{\mu}m}$}
}
\startdata
D2 & 00335$-$2732 & 0:36:00.37 & $-$27:15:33.2 & Interacting group & 0.069 & 
11.78  \\
D5 & 01031$-$2255 & 1:05:36.53 & $-$22:39:18.7 & Interacting group & 0.187 & 
11.87  \\
E1 & 01355$-$1814 & 1:37:57.41 & $-$17:59:20.1 & Interacting group & 0.192 & 
12.20  \\
73 & 06268+3509 & 6:30:13.22 & +35:07:51.1 & Interacting group & 0.170 & 11.92 
\\
B4 & 06487+2208 & 6:51:45.73 & +22:04:28.4 & Multiple nuclei & 0.144 & 12.12 \\
B7 & 08344+5105 & 8:38:03.58 & +50:55:09.9 & Multiple nuclei & 0.097 & 11.76 \\
C3 & 11087+5351 & 11:11:36.36 & +53:34:59.8 & Multiple nuclei & 0.143 & 11.79 \\
52 & 12450+3401 & 12:47:31.75 & +33:44:34.6 & Interacting group & 0.159 & 11.85
\\
25 & 13342+3932 & 13:36:24.14 & +39:17:32.8 & Interacting group & 0.180 & 12.05
\\
27 & 13539+2920 & 13:56:09.93 & +29:05:36.1 & Interacting group & 0.108 & 11.81
\\
85 & 14337$-$4134 & 14:36:59.18 & $-$41:47:06.2 & Multiple nuclei & 0.182 & 
11.80  \\
54 & 16007+3743 & 16:02:32.74 & +37:34:52.9 & Interacting group & 0.185 & 11.53
\\
44 & 18580+6527 & 18:58:13.70 & +65:31:26.1 & Multiple nuclei & 0.176 & 11.90 \\
15 & 19297$-$0406 & 19:32:22.27 & $-$4:00:01.8 & Multiple nuclei & 0.086 & 12.20
\\
17 & 20100$-$4156 & 20:13:29.51 & $-$41:47:35.0 & Interacting group & 0.130 & 
12.44  \\
93 & 20253$-$3757 & 20:28:37.40 & $-$37:47:11.4 & Multiple nuclei & 0.180 & 
11.86  \\
F4 & 22509$-$0040 & 22:53:32.94 & $-$0:24:42.4 & Multiple nuclei & 0.058 & 11.71
\\
F5 & 22546$-$2637 & 22:57:24.37 & $-$26:21:16.6 & Interacting group & 0.164 & 
11.79  \\
56 & 23365+3604 & 23:39:01.33 & +36:21:08.5 & Multiple nuclei & 0.064 & 11.97 \\
F9 & 23515$-$2421 & 23:54:10.38 & $-$24:04:24.8 & Multiple nuclei & 0.153 & 
11.77  \\
\enddata
\end{deluxetable}

\end{document}